\documentclass[prd,nofootinbib,twocolumn,superscriptaddress,preprintnumbers]{revtex4-1}
\usepackage[utf8]{inputenc}
\usepackage{amsmath}
\usepackage{graphicx}
\usepackage{bm}
\usepackage{color}
\usepackage{xcolor}
\usepackage{placeins}

\usepackage[colorlinks]{hyperref}
\hypersetup{
     colorlinks   = true,
     citecolor    = red,
	 linkcolor=blue
}

\newcommand{\bfq}{{\bf q}}
\newcommand{\bfr}{{\bf r}}
\newcommand{\bfp}{{\bf p}}
\newcommand{\bfk}{{\bf k}}

\definecolor{cerulean}{rgb}{0., 0.62,0.9}

\begin{document}

\title{Plasmon production from dark matter scattering}
\author{Jonathan Kozaczuk}
\author{Tongyan Lin} 
\affiliation{Department of Physics, University of California, San Diego, CA 92093, USA}
\date{\today}

\begin{abstract}
We present a first calculation of the rate for plasmon production in semiconductors from nuclei recoiling against dark matter. The process is analogous to bremsstrahlung of transverse photon modes, but with a longitudinal plasmon mode emitted instead.
For dark matter in the 10 MeV -- 1 GeV mass range, we find that the plasmon bremsstrahlung rate is $4-5$ orders of magnitude smaller than that for elastic scattering, but $4-5$ orders of magnitude larger than the transverse bremsstrahlung rate. Because the plasmon  can decay into  electronic excitations and has characteristic energy given by the plasma frequency $\omega_p$, with $\omega_p \approx 16$~eV in Si crystals, plasmon production provides a new signature and method to detect nuclear recoils from sub-GeV dark matter.
\end{abstract}

\maketitle

\section{Introduction}

There have been significant efforts recently to directly detect dark matter (DM) in the low-mass (sub-GeV) regime~\cite{Battaglieri:2017aum}. 
As experiments lower their energy thresholds, collective many-body effects can become increasingly important and enhance the discovery potential beyond that of traditional searches for hard nuclear recoils. Examples can be found in numerous theoretical studies of direct detection of sub-GeV dark matter, including with semiconductors~\cite{Essig:2011nj,Essig:2015cda,Derenzo:2016fse,Griffin:2019mvc}, superconductors~\cite{Hochberg:2015pha,Hochberg:2015fth,Hochberg:2016ajh}, Dirac materials~\cite{Hochberg:2016ntt,Hochberg:2017wce,Coskuner:2019odd,Geilhufe:2019ndy,Kim:2020bwm}, phonon excitations in crystals~\cite{Knapen:2017ekk,Griffin:2018bjn,Cox:2019cod,Campbell-Deem:2019hdx} and in superfluid He~\cite{Schutz:2016tid,Knapen:2016cue,Acanfora:2019con}, and others. 

Recently, Ref.~\cite{Kurinsky:2020dpb} has highlighted a number of low-energy residual rates in experiments achieving the lowest thresholds thus far, and points out the relevance of many-body effects for understanding them. 
The rates are comparable in SENSEI~\cite{Abramoff:2019dfb}, CDMS HVeV~\cite{Agnese:2018col}, and EDELWEISS~\cite{Armengaud:2019kfj,Arnaud:2020svb}, though much lower in DAMIC~\cite{Aguilar-Arevalo:2019wdi}. These experiments all rely on solid-state targets, namely Si and Ge semiconductors.

Ref.~\cite{Kurinsky:2020dpb} has proposed that such excesses could be explained as DM exciting plasmons in semiconductors, since no  excesses with corresponding rates have been observed in  noble liquid experiments such as XENON1T~\cite{Aprile:2019xxb} and DarkSide~\cite{Agnes:2018oej}. One of their proposed ideas is the secondary production of plasmons  during DM-nucleus scattering from DM with mass in the 30 MeV -- GeV range. This could in principle match the observed rates if the probability to produce the plasmon is $\sim 10^{-3} -1$. 

In this work, we provide a first estimate of the plasmon production rate from nuclei recoiling against GeV-scale dark matter, focusing on Si and Ge semiconductors. Plasmons in a semiconductor are the collective oscillations of the valence electrons. The key idea we will use is to approximate the plasmon as a longitudinal mode of a degenerate electron gas (i.e.~a metal). This is justified since plasmons carry an energy of $\omega_p \approx 10-20$ eV, which is  much larger than the band gap $\sim$ eV of a semiconductor. 

The process by which a recoiling nucleus can emit a plasmon is similar to the bremsstrahlung emission of transverse photons, which was previously treated in Ref.~\cite{Kouvaris:2016afs}. Here we consider the bremsstrahlung of longitudinal modes:
\begin{equation}\label{eq:process}
    \chi(p) + N \to \chi(p') + N(q_N) + \omega_L(k)
\end{equation}
where $\chi$ is the dark matter, $N(q_N)$ is a nucleus with energy $E_R = q_N^2/(2 m_N)$, and $\omega_L(k)$ is a plasmon mode with 3-momentum $k$ and energy $\omega_L(k)$.  We will focus on dark matter in the 10 MeV--1 GeV mass range. Then the energy scales for the plasmon and nuclear recoils are both $\gtrsim $ eV, larger than the highest phonon energy $\sim 40-60$ meV in a Ge or Si crystal. As a result, we will treat the DM interaction as scattering off of a free ion (nucleus surrounded by tightly-bound core electrons). The recoiling ion is a current source and can lose energy into both transverse photon and longitudinal plasmon modes.

With these approximations, we find that the rate for plasmon production through the process in Eq.~\ref{eq:process} is typically 4-5 orders of magnitude smaller than the elastic nuclear recoil rate, and therefore cannot explain the excesses studied in Ref.~\cite{Kurinsky:2020dpb}. (Note that the mechanism of Ref.~\cite{Kurinsky:2020dpb} involved a plasmon produced in association with many phonons, and is therefore not captured by our approach.) Nevertheless, bremsstrahlung emission of plasmons by a recoiling nucleus is a novel signature of dark matter scattering in semiconductor targets, and we find that the corresponding rate is around 5 orders of magnitude larger than that for bremsstrahlung emission of transverse modes. Because plasmons can be detected in the form of electronic energy, this process can be used to  extend the reach of current experiments to much lower DM masses.

The rest of this study is structured as follows. We will begin in Sec.~\ref{sec:toy} with an introduction to the physics of plasmons and provide an estimate for the plasmon rate in a metal. We then discuss plasmon production in semiconductors in Sec.~\ref{sec:semi}, computing the rate using a classical approach (an alternative quantum mechanical derivation is provided in an appendix). In Sec.~\ref{sec:rate}, we use these results to estimate the potential reach of a plasmon search in Si and Ge, comparing against the sensitivity provided by elastic nuclear recoils and the Migdal effect, wherein an electron is excited in the nuclear recoil~\cite{Ibe:2017yqa}. We conclude in Sec.~\ref{sec:concl}.

\section{Plasmon emission in an electron gas}\label{sec:toy}

To illustrate the essential ideas surrounding plasmon production in materials, we will start with a simplified scenario: the textbook model of a metal. Here, we have a background of heavy ions surrounded by a free degenerate gas of valence electrons. Because the electrons have a fast response time, we can treat the background of ions as fixed when studying the linear response of the system to perturbations. In this setup, there is a collective mode of longitudinal electron oscillations, the plasmon. Poisson's equation in the absence of external charges, $\hat\epsilon_L(\omega,\bfk)\bfk\cdot \mathbf{E}=0$, implies that collective longitudinal oscillations can occur when $\hat\epsilon_L(\omega,\bfk)=0$, where $\hat\epsilon_L(\omega,\bfk)$ is the longitudinal dielectric function of the material. A plasmon mode therefore corresponds to $\hat\epsilon_L(\omega,\bfk)=0$.

To see the presence of this mode, we start with the Lindhard formula for the longitudinal dielectric function in a crystal at zero temperature~\cite{DresselGruner}: 
\begin{align}\label{eq:Lindhard}
    \hat \epsilon_L&(\omega,\bfk) = 1 +  \lim_{\eta \to 0} \frac{4\pi \alpha_{em}}{V |\bfk|^2}  \times  \\
    & \sum_{\bfp} \Big\{ \frac{| \langle \bfp + \bfk | e^{i \bfk \cdot \bfr} | \bfp \rangle|^2}{\omega_{\bfp + \bfk} - \omega_{\bfp} - \omega - i \eta} +    \frac{| \langle \bfp | e^{i \bfk \cdot \bfr} | \bfp - \bfk  \rangle|^2}{\omega_{\bfp - \bfk} -\omega_\bfp + \omega + i \eta} \Big\} \nonumber
\end{align}
where we are summing over all occupied electron Bloch states $|\bfp \rangle$, $\omega_{\bfp}$ is the energy of the state  $|\bfp\rangle$, $V$ is the volume of the system, and  $\alpha_{em}$ is the fine structure constant. (The sum over different bands has been omitted in this formula to simplify the discussion.) This represents virtual electron-hole excitations that modify the propagation of longitudinal electromagnetic fields.  In particular, this dielectric function is related to the longitudinal electromagnetic polarization tensor $\Pi_L(\omega, \bfk)$ by $\hat \epsilon_L(\omega, \bfk) = 1 - \Pi_L(\omega, \bfk)/|\bfk|^2$, and the plasmon corresponds to a pole in the longitudinal propagator (for reviews that elaborate on this, see e.g.~Refs.~\cite{Raffelt:1996wa,Lin:2019uvt}). 

For a degenerate electron gas, Eq.~\ref{eq:Lindhard} can be evaluated with plane-wave states. Taking the Fermi surface to be spherical and summing over states $ | \bfp \rangle $ with $p < p_F$, where $p =  |\bfp|$ and $p_F$ is the Fermi momentum, one finds
\begin{widetext}
\begin{align}
    \hat \epsilon_L(\omega, \bfk) = 1 + \lim_{\eta \to 0} \frac{3 \, \omega_p^2}{k^2 v_F^2} \Bigg\{ \frac{1}{2} + \frac{p_F}{4 k} \left[ 1 - \left( \frac{k}{2 p_F} - \frac{(\omega + i \eta)}{k v_F } \right)^2 \right] \log \left( \frac{ 1 + k/(2p_F) - (\omega + i \eta)/(k v_F)}{ -1 + k/(2p_F) - (\omega + i \eta)/(k v_F)}\right) \nonumber \\ 
     + \frac{p_F}{4 k} \left[ 1 - \left( \frac{k}{2 p_F} + \frac{(\omega + i \eta)}{ k v_F} \right)^2 \right]  \log \left( \frac{ 1 + k/(2p_F) + (\omega + i \eta)/(k v_F)}{ -1 + k/(2p_F) + (\omega + i \eta)/(k v_F)}\right) \Bigg\}. 
     \label{eq:dielectric_metal}
\end{align}
\end{widetext}
In this expression, the plasma frequency is given by
\begin{equation}
    \omega_p^2 = \frac{4 \pi \alpha_{em} n_e}{m_e}
    \label{eq:plasma_freq}
\end{equation}
where $n_e$ is the number density of valence electrons, $m_e$ is the (in-medium) electron mass, and $v_F \sim 10^{-2}$ is the Fermi velocity. In this work we use units where $c=1$.

The plasmon appears as a zero in Eq.~\ref{eq:dielectric_metal}, which in the small $k$ limit has the form
\begin{align}
    \hat \epsilon_L(\omega,k) \approx 1 - \frac{\omega_p^2}{\omega^2}\left( 1 + \frac{3}{5} \frac{k^2 v_F^2}{\omega_p^2} +  ... \right) \, .
    \label{eq:dielectric_approx}
\end{align} 
Thus the plasmon mode has frequency $\omega_p$ at $k=0$ and has a weak dispersion with momentum. In Eq.~\ref{eq:dielectric_approx}, we have taken the $\eta \to 0$ limit and there is no imaginary part,  but in general there is a finite width $\Gamma$ or inverse damping time in the material,  which can be accounted for by taking $\omega^2 \to \omega^2 + i \omega \Gamma$ in Eq.~\ref{eq:dielectric_approx}. In the free electron gas model, the plasmon is long-lived at small $k$.
Meanwhile, for $k \gtrsim \omega_p/v_F$,
the plasmon dispersion matches onto kinematically-accessible single electron-hole excitations and thus has a large decay width.  Given this large width, the plasmon is only well-defined for $k \lesssim \omega_p/v_F$ (roughly 2.4 keV in Si or Ge).

Because of the momentum cutoff and high energy for plasmons, it is only kinematically possible for DM to excite a single plasmon  if the DM velocity is high, $v \gtrsim 0.01$~\cite{Kurinsky:2020dpb}. 
However, it is possible for plasmons to be %
produced by DM with typical halo velocities of $v \sim 10^{-3}$ if they are produced in association with another excitation such as a nuclear recoil; this gets around the restrictions of the 2-body kinematics by allowing the recoil to absorb most of the momentum. Another way to view this process is from the point of view of the recoiling ion:
a low-energy ion cannot excite the plasmon while satisfying energy and momentum conservation, but in this case an off-shell ion emits the plasmon.

The rate for DM-nucleus scattering with plasmon emission can be obtained in the electron gas model using the machinery of quantum field theory. The process is simply DM-nucleus scattering accompanied by electromagnetic bremsstrahlung radiation~\cite{Kouvaris:2016afs}, but with an external longitudinal mode. We use the results of Ref.~\cite{Braaten:1993jw}, which obtained simple analytic approximations for the $k$-dependent plasmon pole location and residue. The polarization vector for the longitudinal mode in Coulomb gauge is given by
\begin{align}
    \varepsilon_L^\mu = \sqrt{Z_L(k)} \frac{\omega_L(k)}{k} (1, 0, 0, 0)
    \label{eq:polarization}
\end{align}
with wavefunction renormalization given by
\begin{align}
    Z_L(k) \approx 1 - \frac{3}{5} \frac{k^2 v_F^2}{\omega_p^2}  + ...
    \label{eq:wavefunction}
\end{align}
in the $k \ll \omega_p/v_F$ limit. These results are obtained directly from the in-medium longitudinal polarization tensor as described in Ref.~\cite{Braaten:1993jw}. %

In what follows we will restrict ourselves to the soft photon/plasmon limit, defined here to be when the three-momentum of the photon/plasmon $\bfk$ satisfies $|\bfk| \ll |\bfq_N|$ and $|\bfk \cdot \bfq_N|/m_N \ll \omega_p$, where $\bfq_N$ is the momentum of the recoiling ion. This is a good approximation for DM masses in the range $10$ MeV -- 1 GeV, since the typical momentum transfer is on the order of $|\bfq_N| \sim \mu_{\chi N} v \sim 10\, {\rm keV} \times (m_\chi / 10\, {\rm MeV})$, which is much larger than the plasmon cutoff momentum. We have restricted to DM masses $m_\chi \lesssim $ 1 GeV so that $E_R = |\bfq_N|^2/(2 m_N)$ is not too large compared to the typical binding energies of the core electrons. In this limit, we can treat the ions as point particles of charge $Z_{\rm ion}$ and mass $m_N$.

With these assumptions, the differential cross section for a recoiling ion to emit a plasmon in the soft limit is
\begin{align}
    \frac{d^2\sigma_{\rm plasmon}}{dE_R dk} = \frac{2 Z_{\rm ion}^2 \alpha_{em}}{3 \pi} \frac{Z_L(k) k^2}{\omega_L(k)^3} \frac{E_R}{ m_N} \times \frac{d\sigma}{dE_R}{\Bigg|}_{\rm el} \label{eq:rate_toy}
\end{align}
where $E_R = q_N^2/(2m_N)$ is the nuclear recoil energy and $d\sigma/dE_R|_{\rm el}$ is the differential cross section for elastic DM-nucleus scattering, modified to account for the fact that the DM deposits total energy $E_R + \omega_L(k)$. As we argue in the following section, we expect this expression to provide a reasonable approximation for the rate in simple semiconductors as well, and we will use it to compute the production rates from DM scattering in Sec.~\ref{sec:rate}.

In comparison, the bremsstrahlung rate for transverse photons in the soft limit is
\begin{align}
    \frac{d^2\sigma_\gamma}{dE_R dk} = \frac{4 Z_{\rm ion}^2 \alpha_{em}}{3 \pi} \frac{Z_T(k) k^2}{\omega_T(k)^3} \frac{E_R}{ m_N} \times \frac{d\sigma}{dE_R}{\Bigg|}_{\rm el} \label{eq:rate_toy_transverse} \,
\end{align}
where the transverse modes are well-approximated by a dispersion $\omega_T(k) = \sqrt{\omega_p^2 + k^2}$ and $Z_T(k) \approx 1$. In the limit of $k \gg \omega_p$, the plasmon bremsstrahlung rate is enhanced by a large factor of $Z_L(k) k^3/\omega_L(k)^3$; however, this is partially counteracted by the cutoff in plasmon momentum.
Assuming $Z_{\rm ion} = 4$, $E_R \sim 100$ eV, and allowing for $k$ up to a keV, Eq.~\ref{eq:rate_toy}  indicates that plasmon production will be roughly 4 orders of magnitude smaller than the rate for elastic nuclear scattering. This is still significantly larger than the production rate for transverse modes, which is suppressed relative to the elastic recoil rate by roughly 10 orders of magnitude. While the rate to emit plasmons is small, the plasmon is an electronic excitation peaked around $\omega_p$, which provides a complementary signature for nuclear recoils from light dark matter. In the following section, we discuss how this simplified scenario is modified in semiconductors.

\section{Plasmon emission in semiconductors}\label{sec:semi}

In semiconductors such as Si and Ge, the plasmon energy at zero momentum is well-approximated by the plasma frequency $\omega_p$, taking $n_e$ to be the number density of valence electrons and $m_e$ to be the effective electron mass in the material~\cite{PhilippEhrenreich1963}. As discussed above, the plasmon is a zero in the dielectric function or a pole in the longitudinal propagator for electromagnetic fields.
In what follows, we will use classical arguments to derive general results for the energy transfer to soft plasmon and photon modes in terms of the dielectric function. Given experimental data or first-principles calculations for $\hat\epsilon(\omega,\bfk)$, we can in principle account for the many-body physics of a semiconductor. 

We begin this section with a discussion of how the dielectric function in semiconductors differs from that of the simple model in the previous section. 
The first difference appears in the presence of a band gap, $\omega_g \approx 1$ eV. However, for the materials under consideration such as Si and Ge, the plasmon frequency $\omega_p \approx 10-20$ eV is much larger than the band gap $\omega_g \approx$ eV and the corresponding effect is small. This can be seen for example in the Fr\"{o}hlich oscillator model for $\hat \epsilon_L(\omega)$ in semiconductors considered by Refs.~\cite{Kurinsky:2020dpb,KundmanLBL}, which predicts a dielectric function nearly identical to Eq.~\ref{eq:dielectric_approx} for $\omega$ near $\omega_p$ (we discuss this further below). 

In contrast to the electron gas, the band structure of a semiconductor also allows for interband electronic transitions. These contribute to both the real and imaginary parts of $\hat \epsilon_L(\omega,\bfk)$ (see e.g.~Ref.~\cite{DresselGruner}). 
In addition, one needs to account for the electron wavefunctions, which are not described by plane waves. Taking all this into account, we expect the residue of the plasmon pole, the plasmon dispersion relation and width to be sensitive to the band structure and wavefunctions of the electron-hole pairs that contribute to the correlation function. All of this information is encapsulated inside $\hat \epsilon_L(\omega, \bfk)$.

Despite the differences between semiconductors and metals, experimental data suggests that in relatively simple semiconductors, a slight modification of the free electron gas model of Sec.~\ref{sec:toy} can provide a good description of the plasmon pole.  The energy loss by charged particles in a material is characterized by Im($-1/\hat \epsilon_L(\omega, \bfk))$, and the plasmon appears as a pole in this quantity. 
As discussed in Refs.~\cite{Kurinsky:2020dpb,KundmanLBL},  the Fr\"{o}hlich oscillator model describes the plasmon line shape in the $k \to 0$ limit:
\begin{equation}\label{eq:Frohlich}
\operatorname{Im}\left(\frac{-1}{\hat\epsilon_L(\omega, 0)}\right)\simeq \frac{1}{\epsilon_c} \frac{\left(\omega_p^2-\omega_g^2\right) \omega \Gamma}{\left(\omega^2-\omega_p^2\right)^2+\omega^2\Gamma^2}
\end{equation}
where we have identified the quantity $E_p^\prime$ in Ref.~\cite{KundmanLBL} as the effective plasma frequency $\omega_p$, $\epsilon_c $ is the contribution to the dielectric constant from core electrons ($\approx 1$ in Si~\cite{KundmanLBL}) and $\omega_g\sim \mathcal{O}(1)$ eV is an average band gap energy.  For $\omega_g\ll \omega_p$, Eq.~\ref{eq:Frohlich} reduces to the prediction of the Drude-Sommerfeld model of a metal~\cite{DresselGruner}; this is just the free electron gas model of Sec.~\ref{sec:toy}, modified to include a phenomenological relaxation time $\tau=1/\Gamma$ for electronic excitations, as discussed below Eq.~\ref{eq:dielectric_approx}.

\begin{figure}[t]\centering
\includegraphics[width=0.485\textwidth]{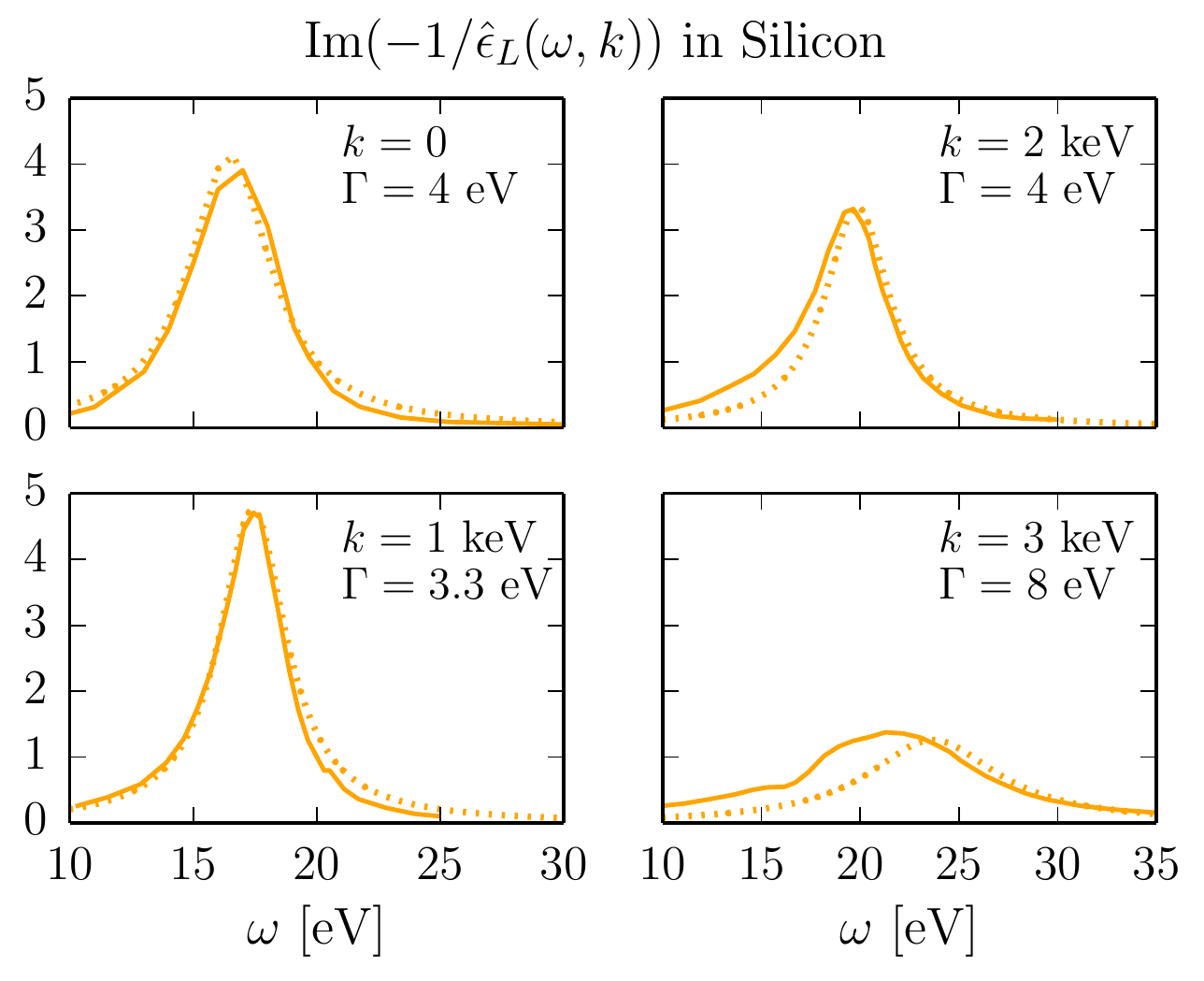}
\caption{\label{fig:Si_Invepsilon}   The energy loss rate for a charged particle into longitudinal modes of a material goes as  Im($-1/\hat\epsilon_L(\omega,\bfk)$), plotted here using collected experimental data on Si (solid lines). Data for $k \to 0$ comes from optical measurements~\cite{opticalconstantsSi} while data for the other $k$ values is from scattering measurements~\cite{SiStructureFactor} (shown here for $\bfk$ along the [111] direction in the crystal). 
The plasmon appears as a zero in the real part of the dielectric function, or as a pole in Im($-1/\hat\epsilon_L(\omega,k)$). The data is well approximated near the pole by the simplified model of Sec.~\ref{sec:toy}: the dotted curves are Eq.~\ref{eq:Frohlich}, adapted with the residue factor $Z_L(k)$ and $\omega_p \to \omega_L(k)$ from Sec.~\ref{sec:toy}. The plasmon width $\Gamma$ is adjusted for each panel.}
\end{figure}

In Fig.~\ref{fig:Si_Invepsilon}, we show Im$(-1/\hat \epsilon_L(\omega, \bfk))$ for Si determined from experimental data.  
For $k=0$, the plasmon is indeed well-described by Eq.~\ref{eq:Frohlich} with $\omega_g\rightarrow 0$, $\omega_p = 16.6$ eV and width of $\Gamma = 4$ eV, as shown in the top left panel of Fig.~\ref{fig:Si_Invepsilon}. For finite $k$, the simplified model of Sec.~\ref{sec:toy} suggests that Eq.~\ref{eq:Frohlich} should be modified to include the residue factor $Z_L(k)$ and $\omega_p\rightarrow \omega_L(k)$. The comparison of the resulting expression to experimental data is shown in the top right and bottom panels of Fig.~\ref{fig:Si_Invepsilon}. Again we find good agreement, especially for smaller $k$ values below $\omega_p/v_F \approx 2.4$ keV (although the width requires some adjusting for each $k$). We therefore expect the results of Sec.~\ref{sec:toy} to provide a reasonable estimate of the plasmon rate once the pole is integrated over. Nevertheless, in what follows we will provide expressions for the energy loss rate for general $\hat \epsilon(\omega,\bfk)$ that can be used away from the plasmon pole and explicitly show how they reduce to the results of Sec~\ref{sec:toy}.

To proceed, we calculate the rate to produce plasmons using classical electrodynamics in a medium. An alternative quantum mechanical derivation is provided in the Appendix. As before, we will make use of the soft photon/plasmon limit.
We assume that DM scatters off one of the nuclei in the material, imparting kinetic energy $E_R$ to the nucleus and the bound electrons. This generates an effective current density
\begin{equation} \label{eq:Jsource}
\mathbf{J}_{\rm ion}=Z_{\rm ion}\,e \,\mathbf{v}_{\rm ion}\,\Theta(t)\,\delta^3(\mathbf{x}-\mathbf{v}_{\rm ion}t)  \end{equation}
where $Z_{\rm ion}$ is equal to the number of valence electrons, $e = \sqrt{4 \pi \alpha_{em}}$ is the unit charge in Heaviside-Lorentz units, and $\mathbf{v}_{\rm ion}$ is the resulting velocity of the scattered ion in the material. Here we neglect the effects of energy loss and damping on the kinetic energy of the ion,  which we assume to be small on the short time scale associated with plasmon production, $t\lesssim 1/\omega_p$. We can also neglect the effects of the ion harmonic potential, since the potential energy of the ion displacement on that time scale is small compared to $E_R$. %
The plasmon will arise as longitudinal $\mathbf{E}$ field oscillations induced by the current in Eq.~\ref{eq:Jsource} and the corresponding response in the material.

Going to Fourier space, one finds the total energy transfer to the material to be 
\small
\begin{equation}
\label{eq:W}
W =  -\int d^3k \int _0^{\infty}  \frac{d\omega}{(2\pi)^4}2\operatorname{Re}\left[\mathbf{J}^*_{{\rm ion}}(\omega,\mathbf{k})\cdot \mathbf{E}(\omega,\mathbf{k})\right]. 
\end{equation}
\normalsize
Focusing on the energy transferred to longitudinal modes, we define the projection $J_{{\rm ion}, L}(\omega, \mathbf{k})=\mathbf{J}_{{\rm ion}}\cdot \mathbf{k}/k$ and similarly for $E_L$. In the soft plasmon limit, $\mathbf{k}\cdot \mathbf{v}_{\rm ion} \ll \omega$, and the longitudinal current density corresponding to Eq.~\ref{eq:Jsource} becomes
\begin{equation}\label{eq:Jsource_Fourier}
J_{{\rm ion}, L}(\omega,\mathbf{k})\simeq \frac{i}{\omega}Z_{\rm ion} \,e \, \mathbf{v}_{\rm ion}\cdot \frac{\mathbf{k}}{k}
\end{equation}
 where we have dropped a term $\propto \delta(\omega-\mathbf{k}\cdot \mathbf{v}_{\rm ion})$ which will not contribute to the plasmon production rate since $\omega \ge \omega_p$ and $\mathbf{k}\cdot \mathbf{v}_{\rm ion}\ll \omega_p$ for the process of interest. This delta-function term would give the usual contribution to the energy loss rate for fast charged particles such as electrons~\cite{KundmanLBL} or millicharged DM~\cite{Kurinsky:2020dpb}, if we take $\mathbf{v}_{\rm ion}$ to be the velocity of the charged particle and consider velocities $v \gtrsim 10^{-2}$ to match onto the plasmon momentum and energy. Instead, the term we have kept in Eq.~\ref{eq:Jsource_Fourier} corresponds to the bremsstrahlung-like contribution from the acceleration of the ion, not present in the standard electron energy loss spectroscopy (EELS) setting.

The field $E_L$ is related to $J_{{\rm ion}, L}$ through the dielectric function of the material. Treating $\mathbf{J}_{\rm ion}$ as an external current, the Fourier space Maxwell-Amp\`ere equation becomes
\begin{equation} \label{eq:EL_JL}
i\,\omega D_L(\omega,\mathbf{k})=i \,\omega \hat{\epsilon}_L(\omega, \bfk) E_L(\omega,\mathbf{k})=J_{{\rm ion},L}(\omega,\mathbf{k}).
\end{equation}
Substituting Eqs.~\ref{eq:Jsource_Fourier} and~\ref{eq:EL_JL} into Eq.~\ref{eq:W} and performing the angular $k$ integration yields
\small
\begin{equation}\label{eq:Wfull}
\frac{dW_L}{dk}=\int_0^{\infty} d\omega \frac{2Z_{\rm ion}^2 \alpha_{em}}{3\pi^2}\left|\mathbf{v}_{\rm ion}\right|^2\frac{k^2}{\omega^3}\operatorname{Im}\left(\frac{-1}{\hat{\epsilon}_L(\omega, \bfk)}\right).
\end{equation}
\normalsize
As expected, the plasmon appears as a pole in $\operatorname{Im}\left(-1/{\hat{\epsilon}_L(\omega, \bfk)}\right)$. However, Eq.~\ref{eq:Wfull} also applies \emph{away} from the plasmon pole, and can be used to compute the total energy deposited through longitudinal excitations in the material (in the soft limit); this accounts for the full dielectric structure of the semiconductor without making the electron gas approximation of the previous section\footnote{For comparison with previous studies of DM-induced  electron and phonon  excitations~\cite{Griffin:2018bjn,Trickle:2019nya}, note that the quantity $\operatorname{Im}\left(-1/\hat{\epsilon}_L(\omega, \bfk)\right)$ is related to the dynamic structure factor by $S(\omega,\bfk) = k^2/(4 \pi^2 \alpha_{em} n_e) \operatorname{Im}\left(-1/\hat{\epsilon}_L(\omega, \bfk)\right)$, where $S(\omega,\bfk)$ describes material response to density perturbations~\cite{Mahan,NozieresPines}}. The same quantity $\operatorname{Im}\left(-1/{\hat{\epsilon}_L(\omega,\bfk)}\right)$ characterizes energy loss by fast electrons in metals or semiconductors~\cite{KundmanLBL,Mahan}.

To make contact with the result of Sec.~\ref{sec:toy}, we approximate  $\operatorname{Im}\left(-1/{\hat{\epsilon}_L(\omega, \bfk)}\right)$ using Eq.~\ref{eq:Frohlich} modified with a factor of $Z_L(k)$ and taking $\omega_p\rightarrow \omega_L(k)$; as noted earlier, this agrees well with the experimentally determined energy loss function in Si (c.f.~Fig.~\ref{fig:Si_Invepsilon}). To isolate the contribution from the plasmon pole, we take the $\Gamma\rightarrow 0$ limit of this expression, which yields
\begin{equation}
\operatorname{Im}\left(\frac{-1}{\hat{\epsilon}_L(\omega,\bfk)}\right)\rightarrow\frac{Z_L(k)\pi \omega_L(k)}{2} \delta\left(\omega-\omega_L(k)\right)
\label{eq:epsdeltafunc}
\end{equation}
for $\omega>0$, where we have used the fact that $\omega_g^2\ll\omega_L(k)^2$ and $\epsilon_c\approx 1$. Noting that the number of plasmons produced at a given energy is $dW_L/\omega$ and performing the $\omega$ integration, we arrive at
\begin{equation}
\frac{dN_{\rm plasmon}}{dk} \simeq \frac{2 Z_{\rm ion}^2 \alpha_{em}}{3 \pi} \frac{ Z_L(k) k^2}{\omega_L(k)^3} \frac{E_R}{ m_N}. \label{eq:Nplasmon}
\end{equation}
This can be interpreted as the probability for producing a plasmon with momentum $k$ for a given nuclear recoil energy, $E_R$. In terms of the cross-section, Eq.~\ref{eq:Nplasmon} corresponds precisely to the prediction of Eq.~\ref{eq:rate_toy}, as anticipated.

A similar calculation can be done for transverse excitations. The current in Eq.~\ref{eq:Jsource} sources a transverse field
\begin{equation}
\mathbf{E}_T(\omega,\mathbf{k})=\frac{i \omega}{k^2-\omega^2 \hat{\epsilon}_T(\omega,\bfk)} \mathbf{J}_T(\omega,\mathbf{k}).
\end{equation}
The corresponding energy loss, $W_T$, is given by the transverse contributions to Eq.~\ref{eq:W}. Noting that the number of photons produced at a given energy is $dW_T/\omega$, the photon production rate is \small
\begin{equation}
\frac{d N_{\gamma}}{d k}=\int d\omega \frac{8 Z_{\rm ion}^2 \alpha_{em}}{3\pi^2}  \frac{E_R k^2}{m_N \omega^2} \operatorname{Im}\left(\frac{-1}{\omega^2 \hat{\epsilon}_T(\omega,\bfk)-k^2} \right).
\label{eq:WTfull}
\end{equation}
\normalsize
In this expression, $\hat\epsilon_T(\omega,\bfk)$ fully characterizes the transverse response of the semiconductor and does not rely on the simplifying assumptions of the model in Sec.~\ref{sec:toy}.

We can again apply the oscillator model to infer an analog of Eq.~\ref{eq:Frohlich} for Im$(-1/(\omega^2\hat\epsilon_T(\omega,\bfk)-k^2))$. Starting from the same Fr\"ohlich model for $\hat\epsilon(\omega,0)$ in e.g.~Ref.~\cite{KundmanLBL}, we compute Im$(-1/(\omega^2\hat\epsilon_T(\omega,0)-k^2))$, identify $k^2+\omega_p^2$ as $\omega_T^2(k)$, and restore an overall residue factor $Z_T(k)$. Then, taking $\omega_g^2 \ll \omega_p^2$ and $\epsilon_c \approx 1$, one finds that for $\Gamma \rightarrow 0$
\begin{equation}
\operatorname{Im}\left(\frac{-1}{\omega^2 \hat{\epsilon}_T(\omega,\bfk)-k^2} \right)\rightarrow \frac{Z_T(k)\,\pi }{2 \, \omega_T(k)} \delta\left(\omega-\omega_T(k)\right).
\end{equation}
Inserting this expression into Eq.~\ref{eq:WTfull} and performing the $\omega$ integration yields the differential probability for exciting a photon with a given $k$. In terms of the production cross-section, the final result matches Eq.~\ref{eq:rate_toy_transverse}.

\begin{figure*}[t!]\centering
\includegraphics[width=0.485\textwidth]{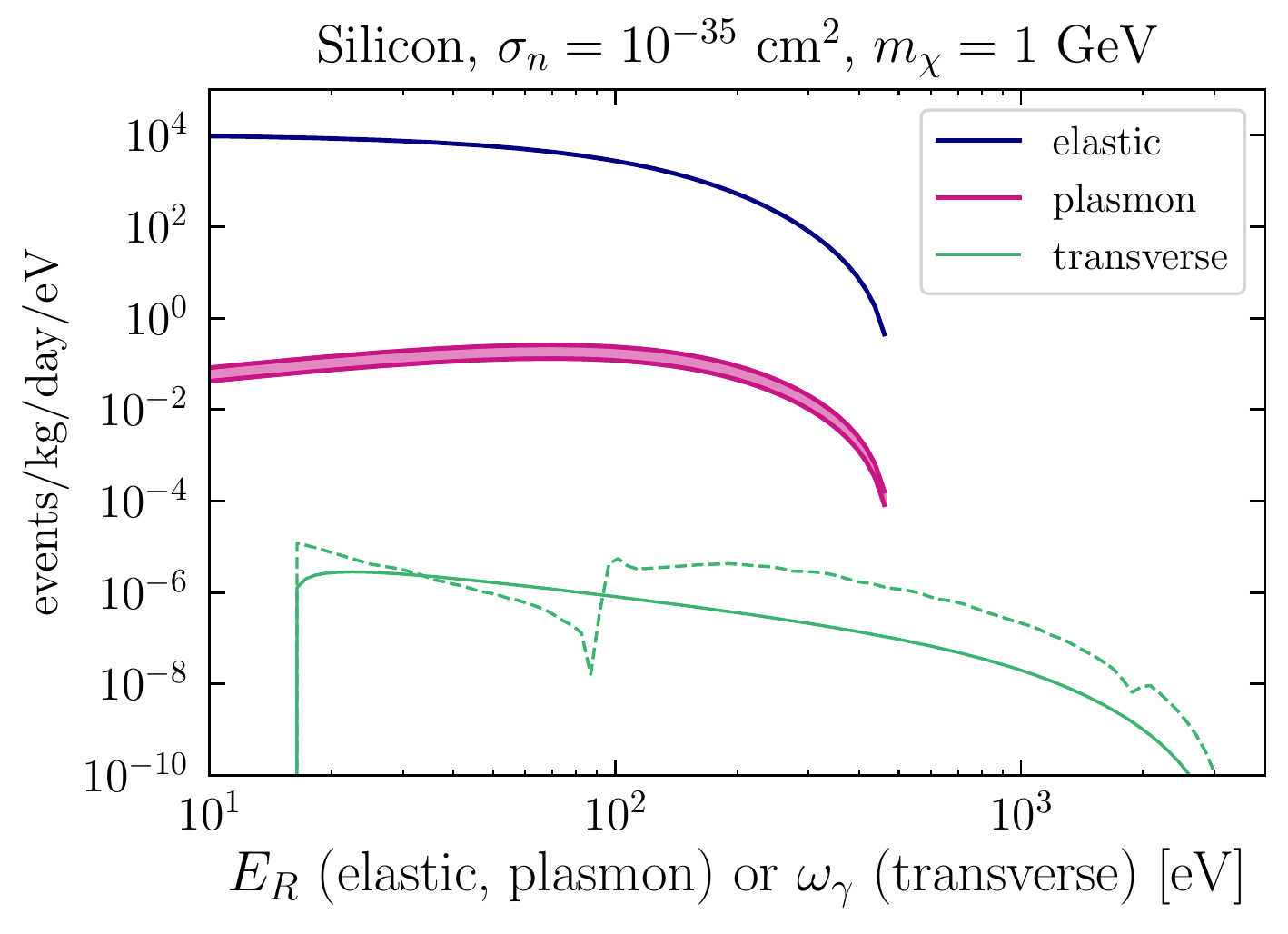}
\includegraphics[width=0.485\textwidth]{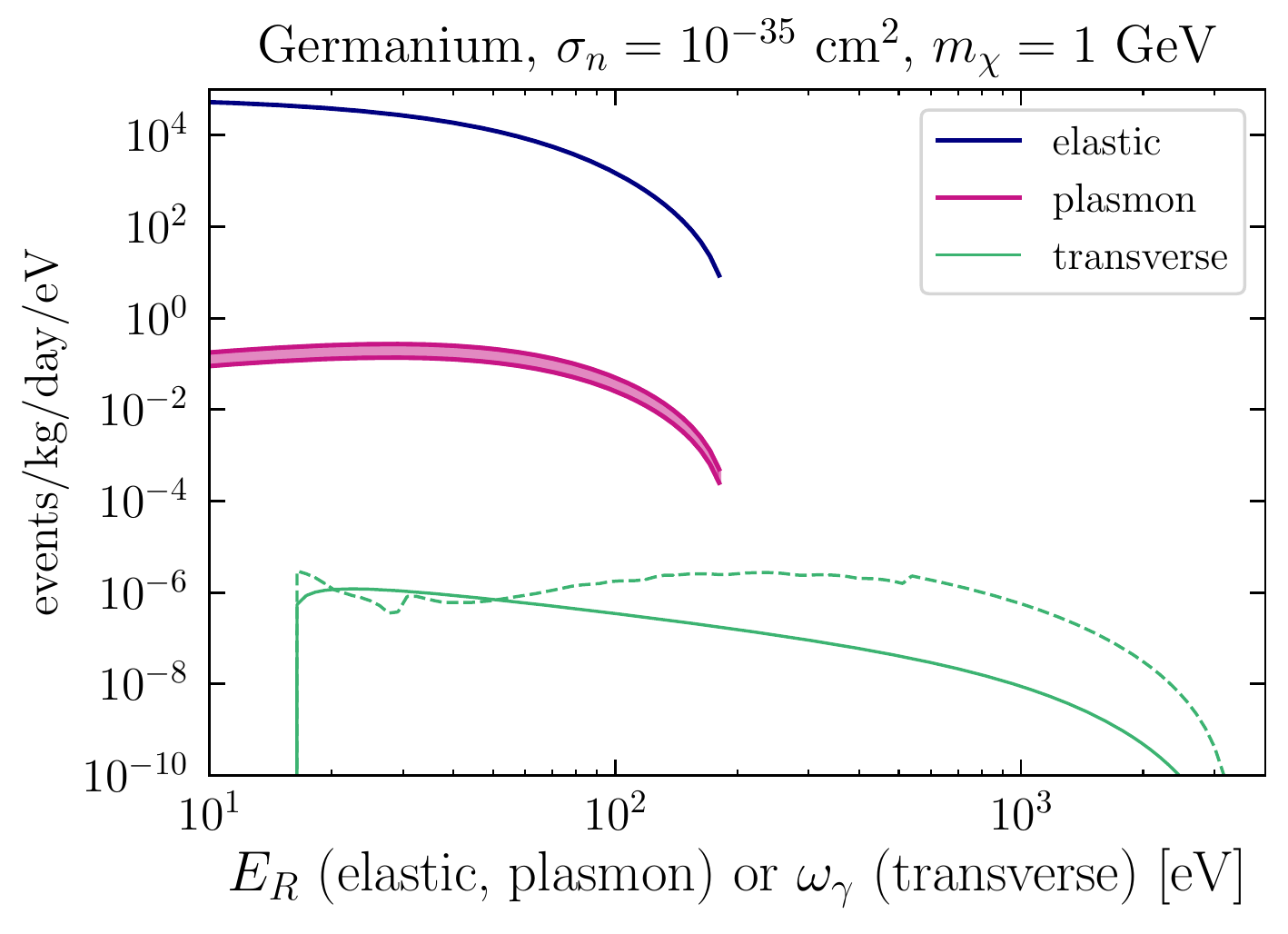}
\caption{\label{fig:rates} Comparison of the differential scattering rate for elastic nuclear recoils and nuclear recoils with plasmon emission. It is assumed that the DM has a spin-independent contact interaction with equal coupling to all nucleons. The band for plasmon emission shows the range of rates if we vary between maximum plasmon momentum of $k_{\rm max} = \omega_p/v_F$ (lower values) up to $k_{\rm max} = 2\omega_p/v_F$ (upper values). We also show rates for bremsstrahlung of transverse modes as a function of photon energy $\omega_\gamma$; the solid lines are obtained using Eq.~\ref{eq:rate_toy_transverse} and the dashed lines use the results of Ref.~\cite{Kouvaris:2016afs} with data on the dielectric functions from Refs.~\cite{opticalconstantsSi,opticalconstantsGe}. Note for Ge the data is limited and the dashed line is uncertain within a factor of few.
}
\end{figure*}

Eqs.~\ref{eq:Wfull}~and~\ref{eq:WTfull} in principle fully characterize the energy loss to plasmons and transverse modes in semiconductors. In order to obtain accurate predictions for DM experiments, a number of effects must be accounted for in these energy loss functions.
In the calculations above, we have used the macroscopic Maxwell's equations and neglected the effects of crystal periodicity.  The relationship between microscopic calculations of $\hat \epsilon(\omega,\bfk)$ and the energy loss functions is modified when taking into account the variation of the microscopic fields over a unit cell; these corrections are often referred to as local field effects~\cite{Adler1962,Wiser1963}. They have been shown to modify the plasmon lineshape and give a better match to electron energy loss spectroscopy data in Si~\cite{Louie1975}. 

In addition, aside from exciting a photon or plasmon, an electron could also be excited above the band gap. In the energy loss rates, this corresponds to a possible continuum of electron recoils away from the plasmon and photon poles. This is similar to the Migdal effect in atoms~\cite{Ibe:2017yqa,Dolan:2017xbu,Bell:2019egg,Baxter:2019pnz}, where electron excitations are created from nuclear recoils; a first approximation for semiconductors was studied in Ref.~\cite{Essig:2019xkx}. Accounting for this effect would again require experimental data or first-principles calculations of the structure factor or dielectric functions.

Besides the plasmon production rate, one must also determine the plasmon decay products, which would ultimately be detected experimentally. The imaginary part of the dielectric function determines the plasmon decay width, where $\Gamma = \omega_p \, {\rm Im}(\hat \epsilon_L(\omega_p,0))$ in the $k\to0$ limit. To infer its decay products, note that the quantity ${\rm Im}(\hat \epsilon_L(\omega,0))$ is closely related to the photoabsorption rate $\sigma_1(\omega) = \omega \, {\rm Im}(\hat \epsilon_L(\omega,0))$; for $\omega$ larger than the band gap, it is dominated by electronic transitions\footnote{In the proposal of Ref.~\cite{Kurinsky:2020dpb}, the plasmon decays dominantly to phonons. Here we attribute the plasmon width and imaginary part of the dielectric function to single electron transitions~\cite{Raether1980}, which is also assumed in studies of bosonic DM absorption at these energies and in the zero momentum limit~\cite{Hochberg:2016sqx,Bloch:2016sjj,Aguilar-Arevalo:2016zop,Abramoff:2019dfb,Arnaud:2020svb}.}. Analogous to the electron gas case, where there is a large plasmon width to single electron excitations for $k \gtrsim \omega_p/v_F$, in semiconductors the plasmon width at zero momentum can be attributed to the availability of electronic transitions with $\omega = \omega_p$~\cite{Raether1980}.  We thus expect that plasmon production leads to energy deposition into electron-hole excitations peaked near $\omega_p$. We will use this fact in the next section when estimating the experimental sensitivity to plasmon production from DM scattering.

\section{Rate results}\label{sec:rate}

We now compute the plasmon production rate from DM-nucleus scattering. Given our assumptions, the total rate to emit plasmons via bremsstrahlung is
\begin{align}
    \frac{dR}{dE_R} = N_T \frac{\rho_\chi}{m_\chi} \int_{v_{\rm min}} d^3 {\bf v} \, v \, f({\bf v}) \int_0^{k_{\rm max}} dk \frac{d^2 \sigma }{dE_R \, dk}.
\end{align}
Here, $N_T$ is the target number density, $\rho_\chi = 0.4$ GeV/cm$^3$ is the local dark matter density, and $f({\bf v})$ is the DM velocity distribution in the Earth's frame, which we take to be the Standard Halo Model with $v_0 = 220$ km/s, $v_e = 240$ km/s, and $v_{esc} = 550$ km/s. Since we are working in the soft limit, we approximate the threshold velocity for exciting a plasmon as
\begin{align}
    v_{min} = \frac{1}{\sqrt{2 m_N E_R}} \left( \frac{m_N E_R}{\mu_{N \chi}} + \omega_p \right)
\end{align}
with $\mu_{N \chi}$ the nucleus-DM reduced mass. This is identical to the threshold velocity for inelastic DM scattering with mass splitting $\delta = \omega_p$. (We have neglected the weak dispersion in the plasmon mode to simplify the velocity integral.)
In order to estimate the effects of the $k$-dependent dispersion and wavefunction renormalization, the rate is computed from Eq.~\ref{eq:rate_toy} using the results of Ref.~\cite{Braaten:1993jw} for $\omega_L(k)$, $Z_L(k)$. As argued in the previous section, this should provide a reasonable estimate of the rate in relatively simple semiconductors.

In Fig.~\ref{fig:rates} we compare the rate for elastic nuclear recoils, bremsstrahlung production of plasmons, and bremsstrahlung production of transverse modes for $m_\chi$ = 1 GeV. Here it is assumed that DM couples equally to all nucleons with a DM-nucleon cross section of $\sigma_n$. Then the elastic scattering cross section is $d\sigma/dE_R|_{\rm el} = A^2 \sigma_n m_N/(2 \mu_{\chi n}^2 v^2)$, where $\mu_{\chi n}$ is the DM-nucleon reduced mass. The nuclear form factor can be neglected for the low energy recoils considered here.

For plasmon emission in both Si and Ge targets we take $\omega_p = 16$ eV~\cite{PhilippEhrenreich1963}. Compared to elastic nuclear recoils, plasmon emission is suppressed by 4-5 orders of magnitude, depending on the maximum plasmon momentum $k_{\rm max}$, which we vary between $\omega_p/v_F$ and $2\omega_p/v_F$. For tranverse bremsstrahlung, we show both the  result derived in our approach, which should be valid for energies below $O(100)$ eV, and the result of Ref.~\cite{Kouvaris:2016afs}, which was computed for atomic targets and thus not appropriate for low energies. We expect the full result to interpolate between these two, but we defer a more detailed analysis of this to future work. In either case, the rate for transverse photon emission is smaller than the plasmon emission rate by another $\sim$5 orders of magnitude, in line with the discussion of Sec.~\ref{sec:toy}. 

\begin{figure*}[t!]\centering
\includegraphics[width=0.485\textwidth]{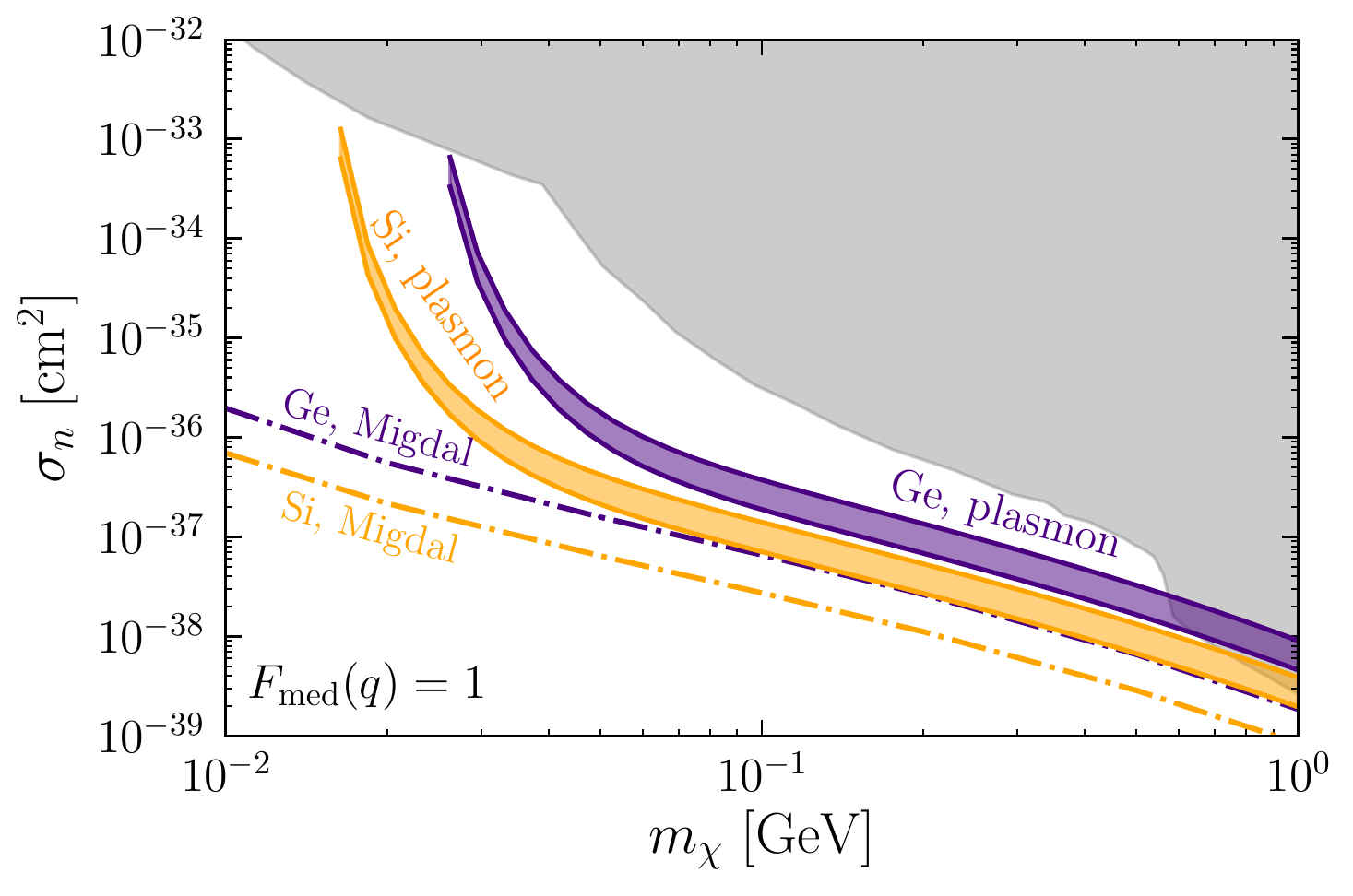}
\includegraphics[width=0.485\textwidth]{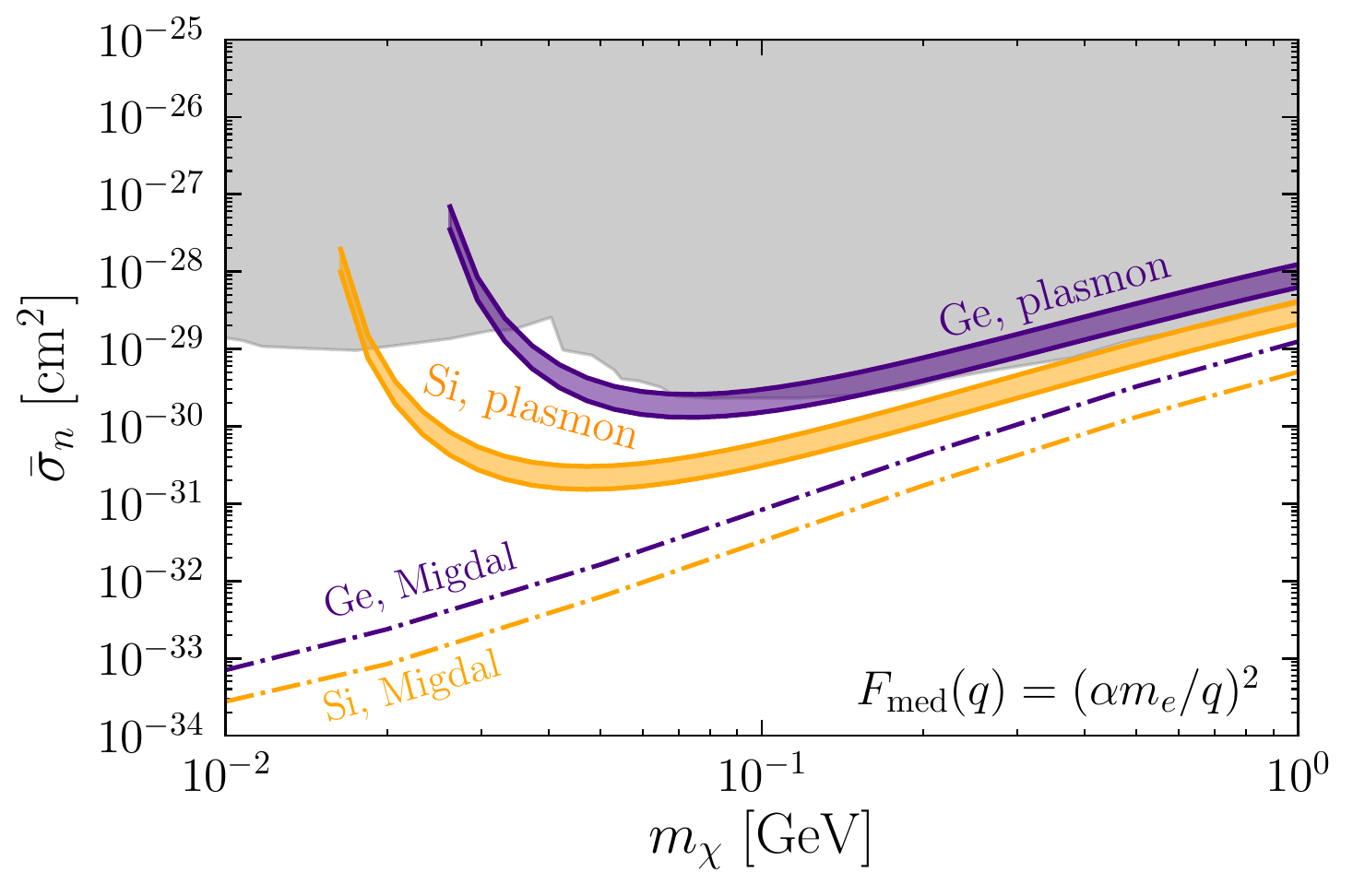}
\caption{\label{fig:sensitivity}  Projected sensitivity to sub-GeV dark matter from plasmon production. Also shown is the sensitivity from estimates of the Migdal rate in semiconductors from Ref.~\cite{Essig:2019xkx}; these may be uncertain within an order of magnitude.
All curves are drawn with kg-year exposure and zero background events. Assuming the plasmon decays to electron excitation with $O(1)$ probability, it will yield on average $\sim 5$ electrons with high efficiency. The differences in behavior of the plasmon and Migdal rates at low masses is due primarily to the choice of threshold $E_R$ (nuclear recoil energy), which we have taken here to be 100 meV to avoid the phonon regime. The gray shaded areas include constraints from XENON1T~\cite{Aprile:2019jmx}, LUX~\cite{Akerib:2018hck}, a recast of XENON10~\cite{Angle:2011th}, XENON100~\cite{Aprile:2016wwo}, and XENON1T~\cite{Aprile:2019xxb} data in terms of the Migdal effect from Ref.~\cite{Essig:2019xkx}, as well as (left plot only) constraints from  CRESST III~\cite{Abdelhameed:2019hmk} and CDEX~\cite{Liu:2019kzq}.
}
\end{figure*}

Plasmon emission is relatively more important for larger DM masses and more energetic ions, which can be seen in the factor of $E_R/m_N$ in the differential cross sections. For $m_\chi < 1$ GeV, the probability for plasmon emission is thus even smaller than that shown in Fig.~\ref{fig:rates}. However, the plasmon can deposit energy in electronic excitations, so this can still be a promising way to search for low-energy nuclear recoils from DM, as we will discuss below.

Finally, while we do expect the probability for plasmon emission to grow for $m_\chi > 1$ GeV, we caution against numerical extrapolation of our results to much higher masses. This is because we have treated the nucleus and core electrons together as a point particle. For heavier DM, there is sufficient energy in nuclear recoils to also ionize core electrons, and the bremsstrahlung rates may be even larger since the screening of the nucleus electric charge is less effective.

\subsection{Sensitivity for low-threshold experiments}

Plasmon production is an additional scattering mechanism that contributes electron excitations or charge signals from nuclear recoils. Not accounting for such charge signals, the typical thresholds for detecting nuclear recoils in current or upcoming experiments is $E_R \gtrsim 30$ eV, corresponding to sensitivity to $m_\chi \gtrsim 0.5$ GeV~\cite{Abdelhameed:2019hmk,Agnese:2016cpb}. On the other hand, plasmon decay to single electron excitations with energy $\approx 16$ eV would yield on average $\sim 5$ measured electrons in Si or Ge~\cite{AligBloomStruck}. This is well above the charge threshold in low-threshold semiconductor experiments such as Refs.~\cite{Abramoff:2019dfb,Agnese:2018col,Arnaud:2020svb,Aguilar-Arevalo:2019wdi}. Thus, nuclear recoils that are not energetic enough to be observed directly can still result in an observable charge yield from plasmon emission.

In Fig.~\ref{fig:sensitivity}, we show the sensitivity to light dark matter from plasmon emission, assuming 100 g-year exposure and zero background. 
Note that for plasmon decays yielding $\sim 5$ electrons on average, it is not necessary  to assume zero background in the 1- or 2-electron bins, where there may in fact be large backgrounds in a realistic experimental analysis. The total plasmon rate is calculated assuming $E_R >$ 100 meV; for lower energies, the ion kinematics assumed here are no longer accurate and single- and multi-phonon~\cite{Campbell-Deem:2019hdx} production will start to dominate. The turnover in the sensitivity curves at around $m_\chi \approx 30$ MeV is due to our choice of threshold $E_R$. The left panel of Fig.~\ref{fig:sensitivity} assumes a contact interaction between the DM and the nucleus, while the right panel shows the massless mediator case, where we have included an additional DM-mediator form factor $F^2_{\rm med}(q) = (\alpha m_e /q)^4$. The sensitivity is noticeably worse for light mediators since plasmon production scales as $E_R/m_N$.

Charge signals for light DM can also be produced through the Migdal effect~\cite{Ibe:2017yqa,Bell:2019egg,Essig:2019xkx,Baxter:2019pnz}, wherein a recoiling nucleus can excite or ionize electrons. While the first discussions of the Migdal effect considered isolated atomic systems, Ref.~\cite{Essig:2019xkx} estimated the corresponding effect in semiconductors, and we show those results in  Fig.~\ref{fig:sensitivity} for comparison. Both plasmon emission and the Migdal effect rates feature a $q_N^2/m_N^2$ suppression, leading to similar behavior in the sensitivity curves. They start to deviate from one another for $m_\chi \lesssim 20-30$ MeV because in Ref.~\cite{Essig:2019xkx} the rates are integrated over all $E_R$ assuming free nuclear recoils, while we have set $E_R > 100$ meV to avoid the phonon regime. In the future, it would be interesting to account for phonon dynamics, and to compare the ionization signals off the plasmon pole to the atomic Migdal effect.

In our reference model, we have assumed that DM has spin-independent contact interactions with all nucleons, such that rates scale as $A^2$. If the DM couples to electrons and protons through a dark photon mediator, all of the rates going through nuclear recoils are smaller by $Z_{\rm ion}^2/A^2$ for the mass range discussed here. In this model, DM-electron scattering would typically provide stronger constraints~\cite{Baxter:2019pnz,Essig:2019xkx}. 

\section{Discussion}\label{sec:concl}

Plasmons can significantly impact several aspects of dark matter production and detection. Dark sector particles can be produced through plasmon decay or conversion in stars~\cite{Raffelt:1996wa,An:2013yfc,An:2013yua,Hardy:2016kme,Chang:2018rso,Chu:2019rok,Mikheev:1998bg,Caputo:2020quz} or in the early universe~\cite{Dvorkin:2019zdi}. Plasmons also play a role in the interaction of charged DM in Galactic dynamics~\cite{Li:2020wyl}. As for direct detection, bosonic dark matter can be absorbed into plasmon modes~\cite{Hochberg:2016sqx,Hochberg:2016ajh,Lawson:2019brd}.

Here, we have taken the first steps towards calculating the plasmon production rate from nuclear recoils in a solid state target material.  We have treated the plasmon mode in the semiconductor as similar to that in a degenerate electron gas. In modeling the initial nuclear recoil, we treated the ion as a recoiling point particle, which loses energy by creating electronic excitations including the plasmon.  
Plasmon production can be further elucidated by taking into account the semiconductor band structure, possible anisotropic crystal structure, and the role of phonons. Detailed studies of the plasmon pole and decay modes in semiconductors, particularly at large momenta, will allow us to obtain more accurate rate calculations. Finally, calibration data with sources demonstrating plasmon production will be needed in order to set precise limits or search for DM with plasmon production.

We have found that the rate for producing plasmons via bremsstrahlung off nuclear recoils is insufficient to explain the direct detection rates highlighted by Ref.~\cite{Kurinsky:2020dpb}. This emission of a plasmon from an off-shell ion occurs on short time scales $\sim 1/\omega_p$ in the nuclear recoil.
Meanwhile, Ref.~\cite{Kurinsky:2020dpb} proposes secondary plasmon production from a nonlinear interaction involving multiphonon production; such a process would have to occur in a different kinematic regime or  on a different timescale than considered here.

Beyond such excesses, plasmon production provides a complementary way to search for nuclear recoils from low-mass DM. The approach is similar in spirit to searches for recoil-associated bremsstrahlung of transverse modes from DM-nucleus scattering and to the Migdal effect. In all of these cases, it is possible to use the electronic energy to improve the detectability of low mass DM, since currently charge thresholds are much lower than thresholds for nuclear recoil energy deposited in the form of heat or scintillation light. Furthermore, our approach accounts for ionization signals away from the plasmon or photon poles. We plan to investigate charge signals off the plasmon pole and derive limits accounting for all of these processes in future work.

\vspace{0.5cm}

{\emph{Acknowledgements}} --
We thank Dan Baxter, Alvaro Chavarria, Michael Fogler, Yoni Kahn, Simon Knapen, Gordan Krnjaic, Noah Kurinsky, Jung-Tsung Li, John McGreevy, Kaixuan Ni, Katelin Schutz, and Tien-Tien Yu for helpful discussions and feedback on this paper.  We also thank Simon Knapen for collaboration on related work, and Tien-Tien Yu for providing rates for the Migdal effect in Si and Ge. This work was supported in part by an Alfred P. Sloan foundation fellowship and the Department of Energy under grant DE-SC0019195.

\appendix

\section{Plasmon production in quantum mechanics}

In this appendix, we provide an alternate derivation of the energy loss rate in Eq.~\ref{eq:Wfull}. We evaluate the rate for the recoiling ion to lose energy to longitudinal electronic excitations using quantum mechanics. We start with the matrix element for $\chi(p_\chi) + N + |\bfp \rangle \to \chi(p_\chi') + N(q_N) + |\bfp'\rangle$, where $|\bfp'\rangle$ is an excited electron state and $|\bfp\rangle$ is an electron in the ground state of the crystal. The matrix element is
\begin{align}
    {\cal M}_{\bfp \to \bfp'} &= {\cal M}_{\rm el} \frac{ 4 \pi \alpha_{em} Z_{\rm ion} }{V k^2 \hat \epsilon_L(\Delta \omega, \bfk)} \times \langle \bfp' | e^{i \bfk \cdot \bfr} | \bfp \rangle \times \\
    &\Big\{ \frac{1}{\Delta \omega - \bfk \cdot \bfq_N/m_N - k^2/2m_N} - \frac{1}{\Delta \omega + k^2/2m_N}  \Big\}. \nonumber
\end{align}
Here ${\cal M}_{\rm el}$ is the matrix element for elastic DM-nucleus scattering, and $\Delta \omega = \omega_{\bfp'} - \omega_{\bfp}$ is the energy difference in the electron states. In addition, for longitudinal excitations of energy deposition $\Delta \omega$ and momentum transfer $\bfk$ we have taken a Coulomb interaction that is screened by $\hat \epsilon_L(\Delta \omega, \bfk)$. For Bloch states, the matrix element $\langle \bfp' | e^{i \bfk \cdot \bfr} | \bfp \rangle$ is only nonzero if $\bfp' = \bfk + \bfp$, up to a reciprocal lattice vector (for simplicity, we set this to zero in the following discussion).

We next expand in the soft limit where $\bfk \cdot \bfq_N/m_N \ll \omega$, sum over all initial and final electron states, and include a factor of unity in the form of $\int d \omega \, \delta(\omega - \Delta \omega)$:
\begin{align}
    \sum_{\bfp, \bfp'} \ & |{\cal M}_{\bfp \to \bfp'}|^2 = \int d\omega \, \frac{|{\cal M}_{\rm el}|^2 }{ \omega^4 } \frac{ 4 \alpha_{em} Z_{\rm ion}^2}{V |\hat \epsilon_L(\omega, \bfk)|^2} \frac{|\hat \bfk \cdot \bfq_N|^2}{m_N^2} \times \nonumber \\
     & \left( \frac{4 \pi^2 \alpha_{em} }{k^2 V} \sum_{\bfp} 
        | \langle \bfp + \bfk | e^{ i \bfk \cdot \bfr} | \bfp \rangle|^2 \, \delta(\omega - \Delta \omega) \right).
\end{align}
The expression in the second line of the above equation can be identified with Im$(\hat \epsilon_L(\omega, \bfk))$~\cite{DresselGruner}. Summing over all $\bfk$ and averaging over angles, we find the total matrix element squared for excitations into electronic states is 
\begin{align}
     \sum_{\bfk, \bfp, \bfp'} |{\cal M}_{\bfp \to \bfp'}|^2 &=  |{\cal M}_{\rm el}|^2 \times \frac{2 Z_{\rm ion}^2 \alpha_{em}}{3\pi^2} |{\bf v}_{\rm ion} |^2 \times \nonumber \\
     & \int dk \int_0^\infty d \omega  \,  \frac{k^2}{\omega^4}  \, {\rm Im}\left( \frac{-1}{\hat \epsilon_L(\omega, \bfk)} \right)  .
\end{align}
The elastic scattering matrix element is multiplied by a factor identical to the energy loss of Eq.~\ref{eq:Wfull}, but with an extra factor of $1/\omega$ here since we are just computing the rate to produce excitations. Integrating over the ion and dark matter phase space, we find that the cross section to produce electronic excitations in association with the nuclear recoil is the same as the result in the main text. 

\bibliography{plasmon.bib}

\end{document}